\documentclass[aps,amsmath,amsfonts,11pt]{revtex4}
\usepackage{graphicx}
\usepackage{epstopdf}
\usepackage{epsfig,graphicx}
\usepackage[english]{babel}
\usepackage{amsfonts}
\usepackage{amsmath}
\usepackage{latexsym}
\usepackage{graphics,bm}
\usepackage{dcolumn}
\usepackage{natbib}
\usepackage{bm}
\usepackage{rotating}

\begin{document}

\title{Far-Infrared frequency mode conversion using bulk acoustic phonon modes}

\author{Surabhi Yadav$^{1}$ and Aranya B Bhattacherjee$^{*1, 2}$ }

\address{$^{1 , 2}$ Department of Physics, Birla Institute of Technology and Science, Pilani, Hyderabad Campus, Telangana-500078, India  \\  $^{2}$Department of Physics, School of Applied Sciences,
Univeristy of Science and Technology, Ri-Bhoi, Meghalaya-793101, India }

\email{ aranyabhuti@hyderabad.bits-pilani.ac.in}

\begin{abstract}
The ability to design, fabricate and control systems which can covert photons with dissimilar frequencies has technological implications in classical as well as quantum communications. Laser heating and thermal mechanical motion in conventional micro/nanoscale optomechanical systems hamper the use of these systems in quantum information processing networks. In contrast, we propose an unconventional system comprising of a bulk quartz crystal placed within a  Fabry-P$e^{'}$rot cavity. The pumping laser is in the far-infrared region. We explore the possibility of efficient mode conversion between two optical modes supported by the system, mediated by the bulk acoustic phonons of the quartz crystal. Unlike the earlier optomechanical systems, the dark mode in our proposed system is not decoupled from the mechanical mode and yet it enables the efficient mode conversion. The novel results found in our study can be used to harness the dark state for quantum state transfer. The proposed system is robust against excessive heating.
\end{abstract}

\maketitle
\section{Introduction}
In recent years, the field of cavity optomechanics, where circulating optical fields couple to a mechanical resonator via radiation pressure, has seen tremendous progress. The state of the mechanical oscillator can be prepared, probed and coherently controlled \citep{Aspelmeyer, quantum}. The coherent manipulation of mechanical degrees of freedom can enable applications ranging from mode conversion \citep{hill, dong, xu}, sensitivity metrology \citep{stowe}, quantum state transfer \citep{wang} to quantum information processing \citep{stannigel,le}. Experiments on optomechanical interactions have led to numerous interesting phenomena like strong optomechanical coupling \citep{S, teufel, verhagen}, optomechanically induced transparency (OMIT) \citep{weis, safavi}. The success of technology attempting to implement quantum information networks depends on our ability to control and utilize high frequency and long-lived acoustic modes \citep{Aspelmeyer} . High frequency long-lived mechanical excitation's can store quantum information for extended period of time even in the presence of decoherence and this property makes them an excellent candidate for high-fidelity quantum state transfer schemes. Phonons with frequencies in the gigahertz range have already enabled entanglement between spatially separated mechanical resonators \citep{lee, ockeloen}, single-phonon level quantum control \citep{cohen, hong}. These micro/nano-scale systems have the disadvantage that more advanced quantum schemes cannot be implemented due to laser heating. This led to a novel experimental demonstration of an alternate strategy for accessing high-frequency acoustic excitation within bulk crystalline (quartz crystal cavity) \citep{kharel}. In order to access the 13 GHz phonons and to explore the optomechanical interactions, phase matched Brillouin interaction between two distinct optical cavity modes based on bulk optomechanics led to a new method being proposed to increase the spectral resolving power of a collinear acousto-optical filters \citep{pustovoit}.

Many of the classical and quantum systems operating across a wide range of energies are not compatible with one another and hence cannot be combined together on a same platform. Thus one require new device strategies for converting photons of dissimilar frequencies to combine and harness their different properties. In the past few years, different optomechanical schemes have been proposed and implemented experimentally \citep{ali, hill, dong, xu, bagci, andrews, bochmann}. In all of these schemes, laser heating and thermal mechanical dissipation are major obstacles for efficient mode conversion.

Given the above, we propose in this paper optomechanical coupling of dissimilar optical modes to high-frequency bulk acoustic mode within a macroscopic quartz crystal \citep{kharel} as an alternative approach to efficient optical mode conversion. Our proposal is based on the optomechanical interaction between the elastic waves within a bulk quartz crystal and two longitudinal optical cavity modes. The interaction between the elastic and optical modes comes into existence due to formation of time-varying photoelastic grating which initiates energy exchange between the optical modes via Bragg scattering. We analyze optical intracavity emission, optical mode conversion efficiency and explain the observed results using the concept of bright and dark modes. We show that the formation of dark mode induces efficient mode conversion from one optical mode to the other. To the best of our knowledge, no similar theoretical research has been carried out to explore coherent wavelength conversion of optical photons using bulk optomechanics. We will be working in the Far-Infrared (F-IR) frequency regime where mode conversion efficiency was found to be maximum using the current strategy.

\section{Proposed Model and Theory}
A simplified schematic description of the opto-mechanical system that we intend to investigate is shown in Fig.1. A  quartz crystal at cryogenic temperature ($\approx$ 8 K) [also called the bulk acoustic wave resonator] is placed inside an optical Fabry-P$e^{'}$rot resonator. Our proposed model is based on a recent experiment on high-frequency cavity optomechanics using bulk acoustic phonons \citep{kharel}. Macroscopic phonon modes are produced within the bulk quartz crystal and acoustic reflections at the planar surfaces of the crystal leads to the formation of acoustic Fabry-P$e^{'}$rot resonator which supports standing-wave high frequency elastic modes with frequency $\Omega_m > 10$ GHz. 

\begin{figure}[h]
    \centering
    \includegraphics[scale=0.7]{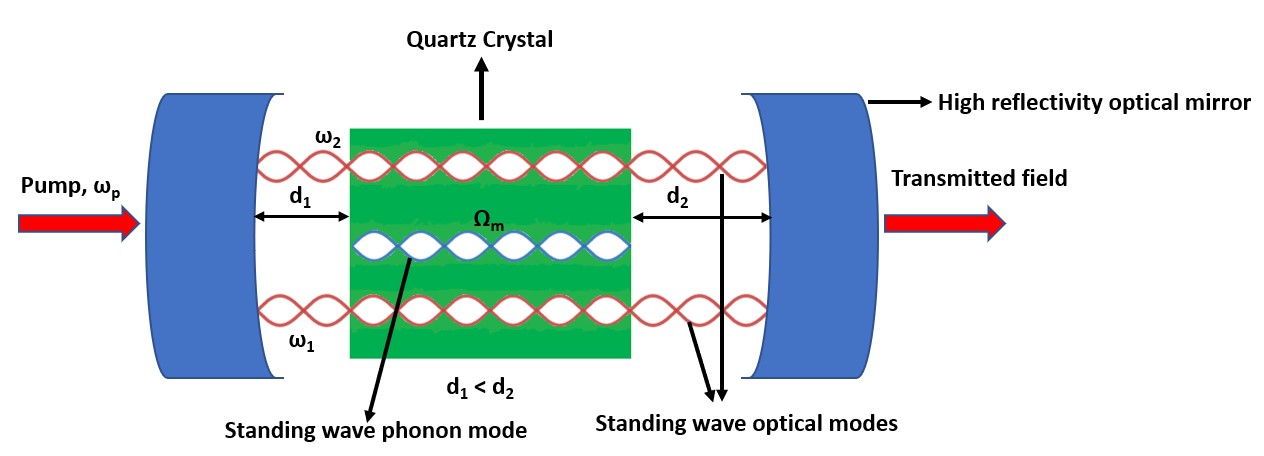}
    \caption{Schematic of an optomechanical system that consists of a bulk acoustic wave resonator that is placed
inside an optical cavity. }
    \label{fig:my_label}
\end{figure}

These high frequency bulk phonon modes acts as a mediator between two longitudinal optical cavity modes. Optomechanical coupling between the acoustic and optical mode occurs when frequency of the time-varying electrostrictive optical force (generated by interference between different optical modes) matches the frequency of the acoustic mode. The motion of the elastic-wave modulates the refractive index of the crystal leading to the formation of a time-varying photoelastic grating. Energy exchange between the different longitudinal optical modes takes place via Bragg scattering by the time-modulated photoelastic grating. In the absence of crystal reflections, equally spaced longitudinal optical modes (uniform density of states) are obtained due to the fact that Bragg scattering rate from the incident laser mode $\omega_j$ to nearest neighbor optical modes $\omega_{j+1}$ and $\omega_{j-1}$ is same. In the presence of optical reflection ($\approx$ 4 $\%$) from the crystal surface, non uniform density of states is obtained \citep{kharel}. This allows one to choose one scattering process over the other (either strokes or anti-strokes scattering). The dynamical Bragg scattering due to the optomechanical interaction within the crystal leads to phase matching $(q_m = k_{j-1}+k_j)$ and energy conservation $(\Omega_m = \omega_j-\omega_{j-1})$. Here $q_m, k_{j-1}, k_j$ are the wave numbers of the acoustic mode, optical mode $\omega_{j-1}$ and $\omega_j$ respectively. These phonon modes which are capable of mediating Bragg scattering lies near the Brillouin frequency $\Omega_B = \Omega_m$ \citep{kharel,boyd}. We will choose a pair of optical modes separated by the Brillouin frequency for our analysis.

The Hamiltonian describing this composite optomechanical system is given by \citep{kharel}
\begin{multline}
    \hat H = \hbar \omega_1\hat a^{\dagger}_1 \hat a_1 + \hbar \omega_2\hat a^{\dagger}_2 \hat a_2 + \hbar \Omega_m\hat b^{\dagger}_m \hat b_m - \hbar g^m_0(a^{\dagger}_2 \hat a_1 \hat b_m + b^{\dagger}_m \hat a^{\dagger}_1 \hat a_2)\\ + \iota\hbar\sqrt{\kappa_1^{ext}}\alpha_p(\hat a^{\dagger}_1 e^{-\iota\omega_p t} +  \hat a_1 e^{\iota\omega_p t}) .
\end{multline}

Here the first two terms of Eqn.(1) are the energy of the two bare optical cavity modes at frequency $\omega_1$ and $\omega_2$. The third term is the energy of the acoustical mode at frequency $\Omega_m$. The two optical modes $\omega_1$ and $\omega_2$ are separated by $\Omega_m$, i.e., $\omega_1$ - $\omega_2$ = $\Omega_m$. The operators $\hat a_1$, $\hat a_2$, and $\hat b_m$ are the annihilation operators of the optical mode 1, mode 2 and the phonon mode respectively. The fourth term denotes the coupling of the two adjacent standing wave optical modes to the $m$-th phonon mode. In the interaction term, $\hat a_2^{\dagger}$ $\hat a_1$ $\hat b_m$ represents the annihilation of the optical mode at frequency $\omega_1$ and a phonon mode at frequency $\Omega_m$ to create an optical mode at a higher frequency $\omega_2$ = $\omega_1$ + $\Omega_m$. This represents the anti-strokes process. The other term  $\hat b_m^{\dagger}$ $\hat a_1^{\dagger}$ $\hat a_2$ represents strokes process. Here $g_0^m$ is the single-photon optomechanical coupling rate given by \citep{kharel}.

\begin{equation}
    g_0^m \approx \frac{\omega_1^2 n^5 p_{13}}{2 c n^2_{eff}}\sqrt{\frac{\hbar}{\rho A L_{ac}\Omega_m}} \frac{L_{ac}}{L_{opt}}  \hspace{2mm} ,
\end{equation}

      \vspace{3mm}
       
where $n_{eff}$ is the effective refractive index of the optical mode, $p_{13}$ is the photoelastic constant of the quartz crystal, $\rho$ is the mass density of the crystal, $L_{ac}$ is the thickness of the crystal, $L_{opt}$ is the spacing between the two cavity mirrors, $A$ is the cross-section area of the crystal, $n$ is the refractive index of the crystal. The last term is the external pump with frequency $\omega_p$ which drives the mode $\omega_1$. Also $\kappa^{ext}_1$ is the loss rate at each cavity mirror and $\sqrt{\kappa^{ext}_1}$ $\alpha_p$ is the rate of pumping the mode 1.

Transforming the Hamiltonian of Eqn.(1) into the rotating frame frequency $\omega_p$ of the pump field, we obtain the following linearized Hamiltonian,
\begin{multline}
    \hat H_{eff} = \hbar \Delta_1 \hat a_1^{\dagger} \hat a_1 + \hbar \Delta_2 \hat a_2^{\dagger} \hat a_2 + \hbar \Omega_m \hat b_m^{\dagger} \hat b_m - \hbar G_m (\hat a_2^{\dagger} \hat a_1 + \hat a_1^{\dagger} \hat a_2) - \hbar G_1 (\hat a_1 \hat b_m + \hat b_m^{\dagger} \hat a_1^{\dagger})\\ - \hbar G_2 (\hat a_2^{\dagger} \hat b_m + \hat b_m^{\dagger} \hat a_2 )+ \iota \hbar \sqrt{\kappa_1^{ext}}\alpha_p(\hat a_1^{\dagger} - \hat a_1) .
\end{multline}

Here, $G_m$ = $g_0^m |b^s_m|$, $G_1$ = $g_0^m |a^s_1|$, $G_2$ = $g_0^m |a^s_2|$. Also, $a_1^s$, $b_m^s$ and $a_2^s$ are the steady state mean values of the operators $\hat b_m$, $\hat a_1$ and $\hat a_2$ respectively. The linearized Hamiltonian can be derived by assuming the optomechanical coupling to be weak and both the cavity modes and the acoustic fields remain near their coherent state \citep{milburn,wilson}. Optical mode conversion mediated by the mechanical resonator in the absence of the interaction $\hbar G_m ((\hat a_2^{\dagger} \hat a_1 + \hat a_1^{\dagger} \hat a_2)$ has been realized experimentally in a system comprising of whispering gallery modes (WGMS) coupled to a mechanical mode in silica resonator \citep{dong}. The Heisenberg equations of motion derived from the Hamiltonian $\hat H_{eff}$ are,

\begin{equation}
   \dot{\hat{a_1}} = -(\iota\Delta_1 + \frac{\kappa_1}{2})\hat a_1 + \iota G_m \hat a_2 + \iota G_1\hat b_m^{\dagger} + \sqrt{\kappa_1^{ext}}\alpha_p \hspace{2mm} ,
   \end{equation}
   
   \begin{equation}
       \dot{\hat{a_2}} = -(\iota\Delta_2 + \frac{\kappa_2}{2})\hat a_2 + \iota G_m \hat a_1  + \iota G_2\hat b_m \hspace{2mm},
   \end{equation}
   
   \begin{equation}
       \dot{\hat{b_m}} = -(\iota\Omega_m + \frac{\gamma_m}{2})\hat b_m + \iota G_1 \hat a_1^{\dagger}  + \iota G_2\hat a_2 \hspace{2mm},
   \end{equation}

\vspace{3mm}

where $\Delta_1$ = $\omega_1$ - $\omega_p$ is the mode 1 - pump field detuning, while $\Delta_2$ = $\omega_2$ - $\omega_p$ is the mode 2-pump field detuning. $\kappa_1$, $\kappa_2$ and $\gamma_m$ are the decay rates of cavity mode 1, mode 2 and phonon mode respectively. Eqn.(4)-(6) will form the basis of our further study related to optical mode conversion in the next section.

\section{Optical Intracavity Emission}
We are interested in the steady state mean response of the system, hence we can replace the operators by their mean expectation values i.e $<\hat a_1>$ = $a_1(t)$, $<\hat a_2>$ = $a_2(t)$ and $<\hat b_m>$ = $b_m(t)$ \citep{kwon,ali}. The mean field steady-state solutions of Eqn.(4)-(6) for the intracavity  steady amplitudes of two optical modes $a_1^s$ and $a_2^s$ are given as, 

\begin{equation}
    |a_1^s|^2 = \frac{A_{1R}^2 + A_{1I}^2}{(l_1^2 + l_2^2 - m_1^2)^2}
\end{equation}

\begin{equation}
    |a_2^s|^2 = \frac{A_{2R}^2 + A_{2I}^2}{[(R_1^2 + R_2^2) - (f_1^2 + f_2^2)]^2}
\end{equation}

\vspace{4mm}

The details of the expression for $|a_1^s|^2$ and $|a_2^s|^2$ are explicitly given in Appendix A. In these expressions, we have introduced the cooperativities of the two optical modes as $C_1$ = $\frac{4 G_1^2}{\gamma_m \kappa_1}$ and $C_2$ = $\frac{4 G_2^2}{\gamma_m \kappa_2}$. Optical emissions from mode 1 and mode 2 are directly proportional to the steady state intracavity intensities $|a_1^s|^2$ and $|a_2^s|^2$ respectively. For demonstration of optical mode conversion in our proposed model, we will use Far Infrared (Far IR, $10\mu$m to 100 $\mu$m wavelength) mode propagating in quartz crystal. For optical applications at tera-hertz frequencies, quartz is an ideal candidate due to its low absorption and high transmission at these frequencies \citep{davies}. In particular, we will be using $\omega_1$ = $2\pi \times 0.99 \times 10^{12}$ Hz and $\omega_2$ = $2\pi \times 10^{12}$ Hz. The frequency $\omega_1$ is excited by the external pump laser and $\omega_2$ is generated by Bragg scattering within the crystal as demonstrated experimentally in the Near-IR region \citep{kharel}. The phonons which mediate the dynamical Bragg scattering within the crystal lies near the Brillouin frequency $\Omega_B$ = $\frac{2 \omega_j n v_a}{v_c}$, where $v_a$ and $v_c$ are the speed of sound and light in the quartz crystal respectively. $n$ is the refractive index of the quartz crystal at frequency $\omega_j$. The two optical modes are separated by the phonon frequency of $\frac{\Omega_m}{2\pi}$ = $90.63$ MHz. Note that the we have used the following experimentally relevant parameters, $n$ = 2.15, $v_a$ = 6327 m/s, $g_0^m$ = $2\pi \times 24$ Hz, $\kappa_1$ = $\kappa_2$ = $2\pi \times 73$ MHz \citep{kharel,davies,roberts,devaty}.

Figure 2(a) and 2(b) shows the normalized emission intensity from mode 1 and mode 2 respectively as a function of cavity cooperativity $C_2$ for the case $\Delta_2/2\pi$ = 65.70 MHz, $\Delta_1/2\pi$ = -25MHz $(\omega_1 - \omega_2 = - \Omega_m)$ and two different values of mechanical damping rate $\gamma_m$ = 0.45 $\kappa$ (dashed plot) and $\gamma_m$ = 0.30 $\kappa$ (solid plot). Such high mechanical damping rate is possible in the Far-Infrared region. To understand the high acoustic damping rate, we can view the propagation of the acoustic field inside the quartz crystal as being similar to the propagation of a laser field in a Fabry-P$e^{'}$rot cavity. The acoustic wavelength corresponding to $\Omega_m$ = 90.68 MHz $(\approx 70 \mu m)$ is much larger than the lateral extent of the excited acoustic mode (beam radius $\approx$ 43 $\mu$m) within the crystal \citep{kharel}. This leads to diffraction loss (analogous to diffraction loss of optical beam in an optical cavity) and the energy in the diffracted acoustic wave which lies outside the acoustic beam radius would be lost. In addition to the intrinsic loss due to diffraction, loss of acoustic energy can also be controlled by adjusting the tilt angle of the crystal axis with respect to the acoustic axis \citep{kharel,ghatak}. As evident from figure 2, emission power from both the modes continuously increases as $C_2$ increases, reaches a peak and then decreases with further increase in $C_2$. The emission power from the modes is higher for a large $\gamma_m$ (= 0.45 $\kappa$) compared to that for a low value of $\gamma_m$ (= 0.30 $\kappa$) but the maximum emission power from mode 2 is higher for a lower $\gamma_m$ as seen in fig.2(b). For a higher $\gamma_m$, the peak in the emission power is reached at lower value of $C_2$.

\begin{figure}[ht]
\hspace{-0.0cm}
\begin{tabular}{cc}
\includegraphics [scale=0.92] {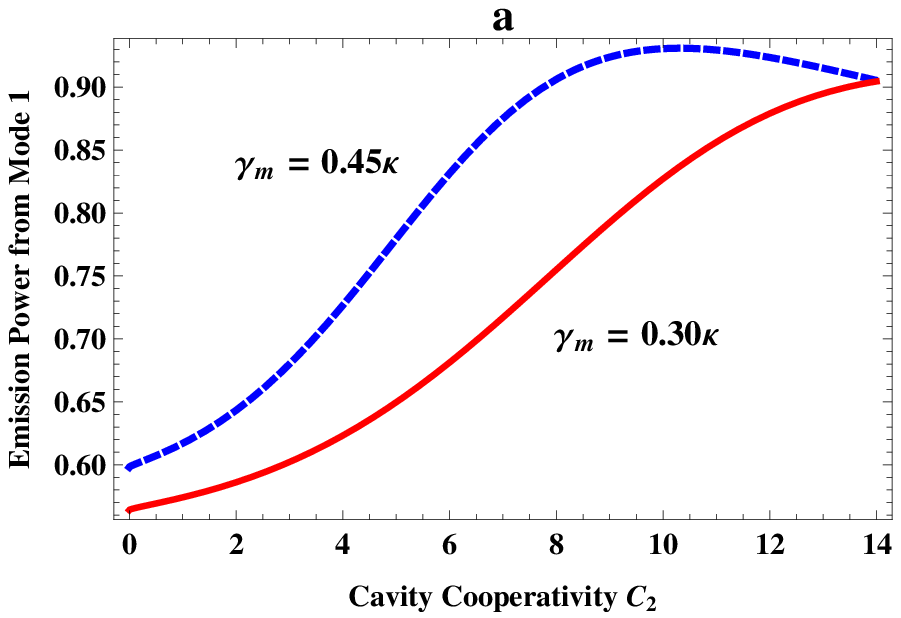} 
\includegraphics [scale=0.9] {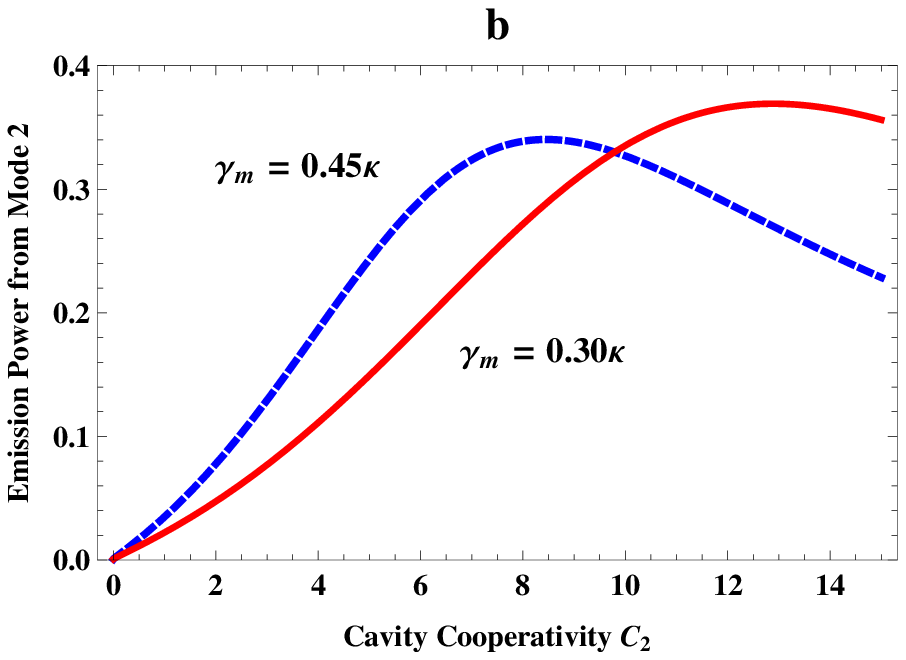}\\
\end{tabular} 
\caption{Normalized optical intracavity emission from mode 1 (fig. 2(a)) and mode 2 (fig. 2(b)) as a function of the cavity cooperativity $C_2$ for two different values of mechanical damping rate $\gamma_m$ = 0.030 $\kappa_1$ (solid line), $\gamma_m$ = 0.45 $\kappa_1$ (dashed line). All system parameters are normalized with respect to the cavity mode 1 damping rate $\kappa_1$. The other system parameters are chosen as $\kappa_2$ = 2 $\kappa_1$, $\Delta_1$ = $-\Omega_m + \Delta_2$, $\Delta_2$ = 0.9 $\kappa_1$, $\Omega_m$ = 1.242 $\kappa_1$, $G_m$ = 0.025 $\kappa_1$, $G_1$ = 0.4 $\kappa_1$.}
\end{figure}

Figure 3(a) and 3(b) shows the plot of emission power from mode 1 and 2 versus $C_2$ for two values of cavity cooperativity $C_1$. $C_1$ = $\frac{4 |g_o^m|^2 |a_1^s|^2 }{\gamma_m \kappa_1}$ can be tuned efficiently by pumping more photons into mode 1 using the external driving laser as observed experimentally \citep{kharel}. As evident from figure 3(a), the increase in emission power from mode 1 as $C_2$ is varied is more for a larger $C_1$. This also leads to an enhanced transfer of energy from mode 1 to mode 2 as $C_1$ is increased from 1.2 to 2.13. This observation is evident from figure 3(b).

\begin{figure}[h]
\hspace{-0.0cm}
\begin{tabular}{cc}
\includegraphics [scale=0.87] {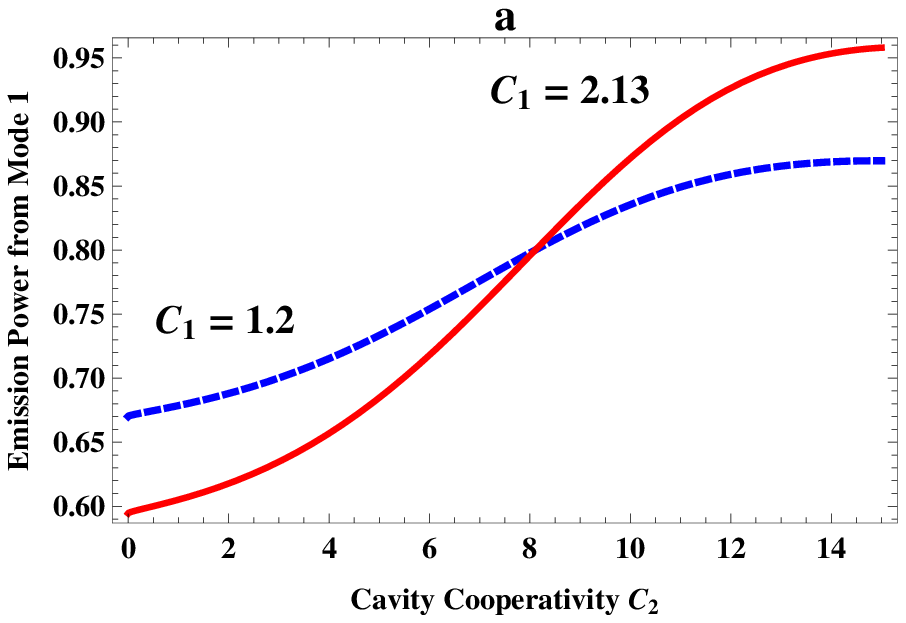} 
\includegraphics [scale=0.87] {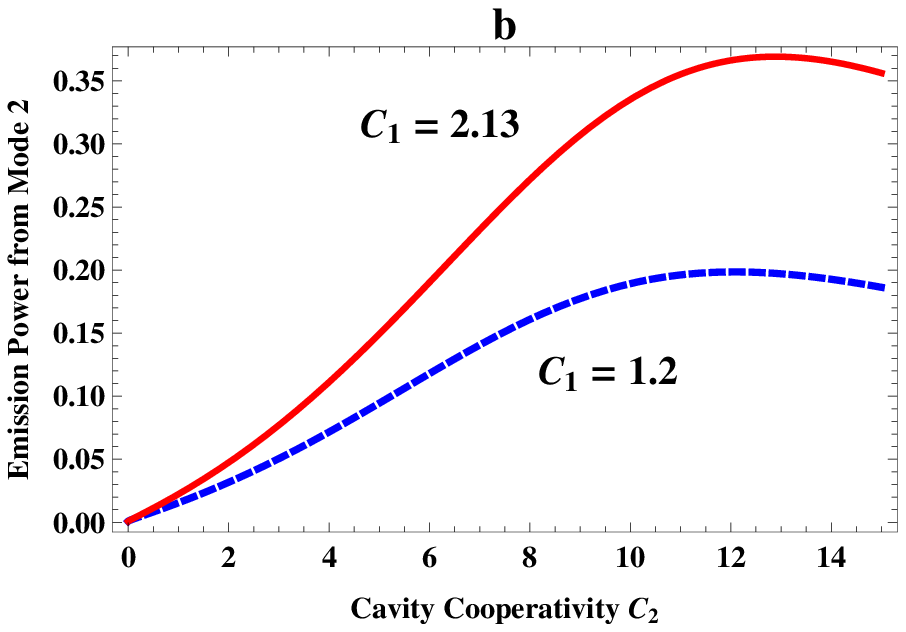}\\
\end{tabular} 
\caption{Normalized optical intracavity emission from mode 1 (fig. 3(a)) and mode 2 (fig. 3(b)) as a function of the cavity cooperativity $C_2$ for two different value of cavity cooperativity $C_1$ = 1.2 (dashed line) and $C_1$ = 2.13 (solid line). The other system parameters are same as that for figure 2 with $\gamma_m$ = 0.3 $\kappa_1$.}
\end{figure}

\section{Optical Mode Conversion Efficiency}
In this section, we analyze the optical mode conversion efficiency ($\eta$) for our proposed model. We define ($\eta$) as the ratio of mode 2 output photon flux over the mode 1 input photon flux $\eta$ = $I_{out}/I_{in}$, $I_{in}$ = $|\alpha_p|^2$ and $I_{out}$ = $\kappa_e^{ext}$ $|a_2^s|^2$.
Thus using results derived in the previous section, the photon mode-conversion efficiency can be written as,

\begin{equation}
    \eta = \frac{\eta_1 \eta_2 \kappa_1\kappa_2 (\tilde{A}_{2R} + \tilde{A}_{2I})}{[(R_1^2 + R_2^2) - (f_1^2 + f_2^2)]^2} \hspace{2mm} ,
\end{equation}

\vspace{3mm}

where $\eta$ = $\frac{\kappa_i^{ext}}{\kappa_i}$ (i = 1,2) is the output coupling ratio for the two optical modes. Here $\tilde{A}_{2R}$ = $\frac{A_{2R}}{\sqrt{\kappa_1^{ext}}\alpha_p}$ and $\tilde{A}_{2I}$ = $\frac{A_{2I}}{\sqrt{\kappa_1^{ext}}\alpha_p}$. The extend of destructive interference in the output of mode 1 determines the efficiency of mode conversion \citep{li}. Figure 4(a) plots the resulting $\eta$ as a function of cavity cooperativity $C_2$ for two values of $\gamma_m$ = 0.45 $\kappa$ (dashed line) and $\gamma_m$ = 0.30 $\kappa$ (solid line). Till $C_2$ = 9, the conversion efficiency is higher for large $\gamma_m$. The peak value of the conversion efficiency corresponding to a lower $\gamma_m$ higher but is attained at a larger value of $C_2$ = 12, which could be difficult to reach with current technology. Fig 4(b) plots $\eta$ versus $C_2$ for two values of $C_1$ = 1.2 (solid line) and $C_1$ = 2.13 (dashed line). Clearly, a higher $C_1$ yields higher mode conversion efficiency.

\begin{figure}[htbp]
\hspace{-0.0cm}
\begin{tabular}{cc}
\includegraphics [scale=0.88] {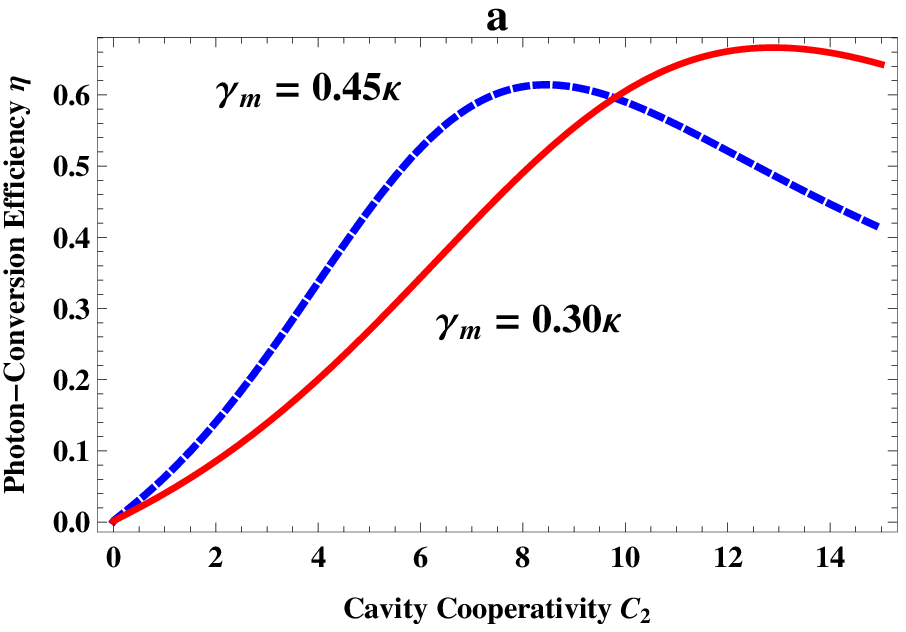} 
\includegraphics [scale=0.87] {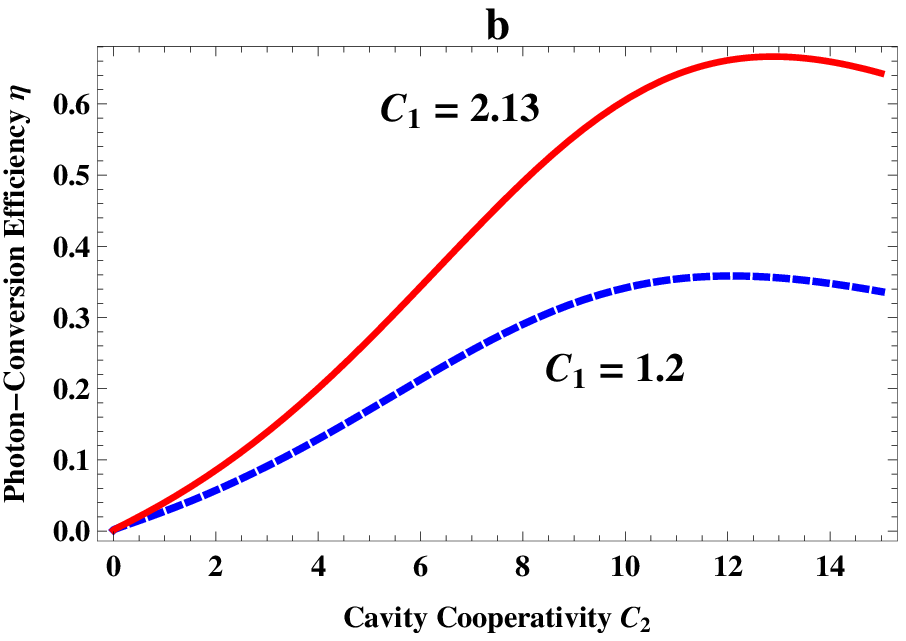}\\
\end{tabular} 
\caption{The photon-conversion efficiency $\eta$ as a function of cavity cooperativity $C_2$ for two different value of $\gamma_m$ (fig. 4(a)) and $C_1$ (fig. 4(b)). The parameters used have the values $\kappa_2$ = 2 $\kappa_1$, $G_m$ = 0.025 $\kappa_1$, $\Omega_m$ = 1.242 $\kappa_1$, $\eta_1 = \eta_2$, $\Delta_2$ = 0.9 $\kappa_1$, $\Delta_2$ = $-\Omega_m + \Delta_2$. For fig. 4(a), $G_1$ = 0.4 $\kappa_1$ and for fig. 4(b), $\gamma_m$ = 1.242 $\kappa_1$.}
\end{figure}

\section{DARK AND BRIGHT MODES}
We now analyze our obtained results using the concept of bright and dark modes.
We now write the system's Hamiltonian in terms of cavity dark and bright modes which are defined as 

\begin{equation}
    \hat a_B = \frac{G_1 \hat a_1 + G_2 \hat a_2 }{\tilde{G}}, \hspace{5mm} \hat a_B = \frac{G_2 \hat a_1 - G_1 \hat a_2 }{\tilde{G}} \hspace{2mm} ,
\end{equation}

\vspace{3mm}

where $\tilde{G_1}$ = $\sqrt{G_1^2 + G_2^2}$. Making use of the above definition of $\hat a_B$ and $\hat a_D$, we rewrite using $\Delta_2 = 0$ and $\Delta_1$ = -$\Omega_m$, the Hamiltonian of eqn. (3) as, 

\begin{multline}
    \hat H_{eff} = -\hbar \Delta_B \hat a_B^{\dagger} \hat a_B - \hbar \Delta_D \hat a_D^{\dagger} \hat a_D + \hbar \Omega_m \hat b_m^{\dagger} \hat b_m - \hbar G_bd (\hat a_B^{\dagger} \hat a_D + \hat a_D^{\dagger} \hat a_B) - \hbar G_{12} (\hat a_D \hat b_m + \hat b_m^{\dagger} \hat a_D^{\dagger} - \hat a_D^{\dagger} \hat b_m + \hat b_m^{\dagger} \hat a_D )\\ - \hbar \tilde{G_1} (\hat a_B^{\dagger} \hat b_m + \hat b_m^{\dagger} \hat a_B^{\dagger} ) - \hbar \tilde{G_2}(\hat a_B^{\dagger}\hat b_m + \hat b_m^{\dagger}\hat a_B)+ \iota \hbar A_1 (\hat a_B^{\dagger} - \hat a_B) + \iota\hbar A_2(a_D^{\dagger} - \hat a_D)
\end{multline}

where, 
\begin{equation*}
    \Delta_D = \frac{G_2^2\Omega_m - 2 G_m G_1 G_2}{\tilde{G^2}}, \hspace{4mm} \Delta_B = \frac{G_1^2\Omega_m + 2 G_m G_1 G_2}{\tilde{G^2}}
\end{equation*}

\begin{equation*}
    G_{bd} = \frac{[G_1 G_2 \Omega_m + G_m + G_m(G_2^2 - G_1^2)]}{\tilde{G^2}}, \hspace{4mm} G_{12} = \frac{G_1 G_2}{\tilde G}
\end{equation*}

\begin{equation*}
  \tilde{G_1} = \frac{G_1^2}{\tilde{G}}, \hspace{4mm} \tilde{G_2} = \frac{G_2^2}{\tilde{G}} 
\end{equation*}

\begin{equation*}
 A_1 = \sqrt{\kappa_1^{ext}}\frac{\alpha_p G_1}{\tilde{G}}, \hspace{4mm}  A_2 = \sqrt{\kappa_1^{ext}}\frac{\alpha_p G_2}{\tilde{G}} 
\end{equation*}

\vspace{2mm}

Equation (11) demonstrates a complex interplay between the two optical modes and the mechanical mode. Using the equations of motion derived from $\hat H^{eff}$, we calculate the steady state values of $a_{B,S}$ and $a_{D,S}$ using $\kappa_1$ = $\kappa_2$ = $\kappa$ . 

 \begin{equation}
    a_{B,S} = A_1(f_{a1} + \iota f_{a2}) + A_2(g_{a1} + \iota g_{a2})
\end{equation}

\begin{equation}
    a_{D,S} = A_1(h_{a1} + \iota h_{a2}) + A_2(J_{a1} + \iota J_{a2})
\end{equation}

The details of the expression appearing in eqns. (12) and (13) is given in Appendix B.

\vspace{5mm}

\begin{figure}[ht]
    \centering
    \includegraphics[scale=0.99] {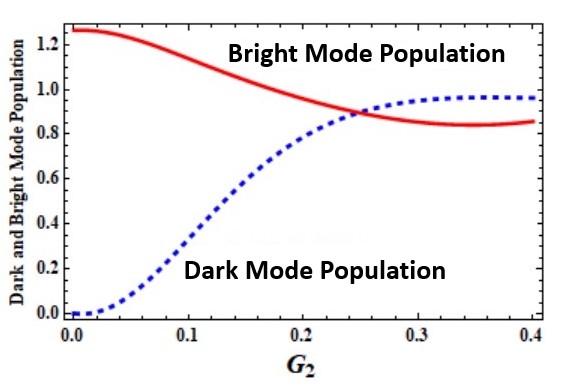}
    \caption{Calculated bright mode and dark mode steady state population as a function of coupling parameter $G_2$. system parameters used are $\kappa_1$ = $\kappa_2$ = $\kappa$, $G_m$ = 0.025 $\kappa$, $\gamma_m$ = 0.2 $\kappa$, $G_1$ = 0.6 $\kappa$ and $\Omega_m$ = 1.242 $\kappa$.}
\end{figure}

In figure 5, we plot the steady state population of dark mode $|a_{D,S}|^2$ and bright mode $|a_{B,S}|^2$ as function of optomechanical coupling strength $G_2$. As $G_2$, increases the cavity cooperativity also increases and as evident from the plot, as $|G_2|$ increases, the bright mode population $|a_{B,S}|^2$ decreases while the dark mode population $|a_{D,S}|^2$ increases. This observation together with the result of previous figures points to the fact that there is a destructive interference between the bright and dark mode and hence optical mode conversion from mode 1 to mode 2 is initiated by suppression of the bright mode \cite{ali}.

\section{Conclusions}
In summary, we have demonstrated efficient frequency conversion between two dissimilar standing-wave longitudinal optical modes in the far-IR region via electrostrictive optical force with a standing-wave longitudinal acoustic mode generated within a bulk quartz crystal. It is clearly shown that the efficiency of mode conversion is better for large cavity cooperativies and mechanical damping rates. In our proposed system, the cavity dark mode, which is a superposition of the two optical modes is not decoupled from the acoustic mode. It is found that when the composite system is driven towards the system's dark mode, the mode conversion efficiency increases. The proposed hybrid quantum system using short-lived phonons within a bulk quartz crystal could be a profitable resource for optical mode conversion.
\cleardoublepage

\section{REFERENCES}

\cleardoublepage

\appendix
\section{Optical Intracavity Emission}
\begin{equation}
    |a_1|^2 = \frac{A_{1R}^2 + A_{1I}^2}{[l_1^2 + l_2^2 - m_1^2]^2}
\end{equation}

\begin{equation}
    A_{1R} = n_1(m_1 + l_1) + n_2l_2
\end{equation}
\begin{equation}
    A_{1I} = n_2(l_1 - m_1) - n_1l_1
\end{equation}

\begin{equation}
    C_1 = \frac{4 G_1^2}{\gamma_m \kappa_1}
\end{equation}

\begin{equation}
    C_2 = \frac{4 G_2^2}{\gamma_m \kappa_2}
\end{equation}

\begin{equation}
    n_1 = A_p(D_1^2 + D_2^2)\frac{\gamma_m}{2}, \hspace{4mm} n_2 = -A_p \Omega_m (D_1^2 + D_2^2)
\end{equation}

\begin{equation}
    m_1 = -G_m \sqrt{C_1}\sqrt{C_2}\frac{\gamma_m \sqrt{\kappa_1 \kappa_2}}{4} [2 D_1\Omega_m + D_2\Omega_m]
\end{equation}

\begin{equation}
    l_1 = D_3 (D_1^2 + D_2^2) + |N_1|^2 D_1 + |N_2|^2 D_1
\end{equation}
\begin{equation}
    l_2 = D_4 (D_1^2 + D_2^2) - |N_1|^2 D_1 + |N_2|^2 D_2
\end{equation}

\begin{equation}
    D_1 = \frac{\kappa_2 \gamma_m}{4} - \Delta_2\Omega_m + \frac{C_2 \gamma_m \kappa_2}{4}, \hspace{4mm} D_2 = \frac{\Delta_2\gamma_m}{2} + \frac{\Omega_m \kappa_2}{2}
\end{equation}

\begin{equation}
    D_3 = \frac{\kappa_1 \gamma_m}{4} + \Delta_1\Omega_m - \frac{C_1 \gamma_m\kappa_1}{4}, \hspace{4mm} D_4= \frac{\Delta_1\gamma_m}{2} - \frac{\kappa_1\Omega_m}{2}
\end{equation}

\begin{equation}
    |N_1|= G_m\Omega_m + \frac{\iota G_m\gamma_m}{2}, \hspace{4mm} |N_2|= G_2 G_1
\end{equation}

\begin{equation}
    |a_2|^2 = \frac{A_{2R}^2 + A_{2I}^2}{[(R_1^2 + R_2^2) - (f_1^2 + f_2^2)]^2}
\end{equation}

\begin{equation}
    A_{2R} = h_1(f_1 + R_1) + h_2(f_2 + R_2)
\end{equation}
\begin{equation}
    A_{2I} = h_1(f_2 - R_2) + h_2(R_1 - f_1)
\end{equation}

\begin{equation}
    h_1 = A_p[D_4 G_m (\Omega_m^2 + \frac{\gamma_m^2}{4}) - N_2(\frac{D_3 \gamma_m}{2} - \Omega_m D_4)]
\end{equation}

\begin{equation}
    h_2 = A_p[D_3 G_m (\Omega_m^2 + \frac{\gamma_m^2}{4}) - N_2(\frac{D_4 \gamma_m}{2} - \Omega_m D_3)]
\end{equation}

\begin{equation}
    f_1 = -2 G_m \sqrt{C_1}\sqrt{C_2} \gamma_m \frac{\sqrt{\kappa_1\kappa_2}}{4} \Omega_m [\Delta_1\Omega_m + \frac{\kappa_1\gamma_m}{4} - \frac{C_1\gamma_m\kappa_1}{4}]
\end{equation}

\begin{equation}
    f_2 = G_m \sqrt{C_1}\sqrt{C_2} \gamma_m^2 \frac{\sqrt{\kappa_1\kappa_2}}{4} [\Delta_1\Omega_m + \frac{\kappa_1\gamma_m}{4} - \frac{C_1\gamma_m\kappa_1}{4}]
\end{equation}

\begin{equation}
    R_1 = D_1[D_3^2 + D_4^2] + D_3[|N_1|^2 + N_2^2]
\end{equation}
\begin{equation}
    R_2 = D_2[D_3^2 + D_4^2] + D_4[N_2^2 - |N_1|^2]
\end{equation}

\section{Dark and Bright Modes}

\begin{equation}
    f_{a1} = \frac{A_{B1}R_9 - A_{B2}R_{10} + A_{B3}R_9 + A_{B4}R_{10}}{(A^2_{B1} + A^2_{B2} - A^2_{B3} - A^2_{B4})}
\end{equation}

\begin{equation}
    f_{a2} = \frac{A_{B1}R_{10} + A_{B2}R_{9} - A_{B3}R_{10} + A_{B4}R_{6}}{(A^2_{B1} + A^2_{B2} - A^2_{B3} - A^2_{B4})}
\end{equation}

\begin{equation}
    g_{a1} = \frac{A_{B1}A_{B5} - A_{B2}A_{B6} + A_{B3}A_{B5} + A_{B4}A_{B6}}{(A^2_{B1} + A^2_{B2} - A^2_{B3} - A^2_{B4})}
\end{equation}
    
\begin{equation}
    g_{a2} = \frac{A_{B2}A_{B5} + A_{B1}A_{B6} - A_{B3}A_{B6} + A_{B4}A_{B5}}{(A^2_{B1} + A^2_{B2} - A^2_{B3} - A^2_{B4})}
\end{equation}

\begin{equation}
   A_{B1} = R_1 - R_5A_{D1} + R_6A_{D2} - R_7A_{D3} - R_8A_{D4} 
\end{equation}

\begin{equation}
   A_{B2} = R_2 + R_5A_{D2} + R_6A_{D1} - R_7A_{D4} + R_8A_{D3} 
\end{equation} 

\begin{equation}
   A_{B3} = R_3 + R_5A_{D3} - R_6A_{D4} + R_7A_{D1} - R_8A_{D2} 
\end{equation}

\begin{equation}
   A_{B4} = R_4 + R_5A_{D4} + R_6A_{D3} - R_7A_{D2} + R_8A_{D1} 
\end{equation}

\begin{equation}
   A_{B5} =  R_5A_{D5} - R_6A_{D6} + R_7A_{D5} - R_8A_{D6} 
\end{equation}

\begin{equation}
   A_{B6} =  R_5A_{D6} - R_6A_{D5} - R_7A_{D6} - R_8A_{D5} 
\end{equation}

\begin{equation}
R_1 = (\Delta_B^2 + \frac{\kappa^2}{4}) (\Omega_m^2 + \frac{\gamma_m^2}{4}) + \Omega_m(\tilde{G_1^2} + \tilde{G_2^2})\Delta_B
\end{equation}

\begin{equation}
R_2 = \frac{\Omega_m\kappa}{2}(\tilde{G_1^2} + \tilde{G_2^2}), \hspace{4mm} R_3 = -2(\tilde{G_1^2} + \tilde{G_2^2}) \Omega_m \Delta_B, \hspace{4mm} R_4 = (\tilde{G_1^2} + \tilde{G_2^2}) \Omega_m\kappa    
\end{equation}

\begin{equation}
    R_5 = -\Delta_B[G_{bd} (\Omega_m^2 + \frac{\gamma_m^2}{4}) + G_{12} \Omega_m (\tilde{G_1^2} + \tilde{G_2^2})]
\end{equation}

\begin{equation}
    R_6 = \frac{\kappa}{2}[G_{bd} (\Omega_m^2 + \frac{\gamma_m^2}{4}) + G_{12} \Omega_m (\tilde{G_1^2} + \tilde{G_2^2})]
\end{equation}

\begin{equation}
  R_7 = - G_{12} \Omega_m \Delta_B (\tilde{G_1^2} + \tilde{G_2^2}) , \hspace{4mm}  R_8 = \frac{G_{12}\Omega_m \kappa}{2} (\tilde{G_1^2} + \tilde{G_2^2})    
\end{equation}

\begin{equation}
    R_9 = \frac{\kappa}{2}(\Omega_m^2 + \frac{\gamma_m^2}{4}), \hspace{4mm} R_{10} = \Delta_B (\Omega_m^2 + \frac{\gamma_m^2}{4})
\end{equation}

\begin{equation}
    A_{D1} = \frac{-\Delta_D (G_{bd} + G_{12} B_R)}{\Delta_D^2 + \frac{\kappa^2}{4}}, \hspace{4mm} A_{D2} =  \frac{\frac{\kappa}{2}(G_{bd} + G_{12} B_R)}{\Delta_D^2 + \frac{\kappa^2}{4}}
\end{equation}  

\begin{equation}
 A_{D3} = \frac{\Delta_D G_{12} B_R}{\Delta_D^2 + \frac{\kappa^2}{4}}, \hspace{4mm} A_{D4} =  \frac{-\frac{\kappa}{2} G_{12} B_R}{\Delta_D^2 + \frac{\kappa^2}{4}}   
\end{equation}

\begin{equation}
    A_{D5} =  \frac{\kappa/2}{\Delta_D^2 + \frac{\kappa^2}{4}}, \hspace{4mm} A_{D6} =  \frac{\Delta_D}{\Delta_D^2 + \frac{\kappa^2}{4}}
\end{equation}

\begin{equation}
    B_R = \frac{\Omega_m (\tilde{G_1^2} + \tilde{G_2^2})}{\Omega_m^2 + \frac{\gamma_m^2}{4} }
\end{equation}

\begin{equation}
    h_{a1} = A_{D1}f_{a1} - A_{D2}f_{a2} + A_{D3}f_{a1} + A_{D4}f_{a2} 
\end{equation}

\begin{equation}
    h_{a2} = A_{D1}f_{a2} + A_{D2}f_{a1} - A_{D3}f_{a2} + A_{D4}f_{a1} 
\end{equation}

\begin{equation}
    J_{a1} = A_{D1}g_{a1} - A_{D2}g_{a2} + A_{D3}g_{a1} + A_{D4}g_{a2} + A_{D5}
\end{equation}

\begin{equation}
    J_{a1} = A_{D1}g_{a1} - A_{D2}g_{a2} + A_{D3}g_{a1} + A_{D4}g_{a2} + A_{D5}
\end{equation}

\begin{equation}
    J_{a2} = A_{D1}g_{a2} + A_{D2}g_{a1} - A_{D3}g_{a2} + A_{D4}g_{a1} + A_{D6}
\end{equation}

\end{document}